\documentstyle[11pt,newpasp,twoside,epsf]{article}
\begin{document}

\def \rstar{\relax \ifmmode \,{\rm R}_\ast \else ${\rm R}_\ast$\fi}
\def \microns{\relax \ifmmode {\, \mu{\rm m}} \else $\mu{\rm m}$ \fi}
\def \mdot{\mbox{$\stackrel{.}{\textstyle M}$}}
\def \teff{\relax \ifmmode {T_{\rm eff}}\else $T_{\rm eff}$ \fi}
\def \asca{{\it ASCA}}
\def \axaf{{\it AXAF}}
\def \astroe{{\it ASTRO-E}}
\def \xmm{{\it XMM}}
\def \Msunyr{\hbox{${\rm M}_\odot\,$  yr $^{-1}$}}
\def \msunyr{\hbox{${\rm M}_\odot\,$  yr $^{-1}$}}
\def \vinf{\relax \ifmmode{v}_\infty \else ${v}_\infty$\fi}
\def \lxlb{\relax \ifmmode {L_{\rm X}/L_{\rm Bol}}\else $L_{\rm X}/L_{\rm Bol}$ 
 \fi}
\def \ecma{\hbox{$\epsilon$ CMa}}
\def \bcma{\hbox{$\beta$ CMa}}
\def \iras{{\it IRAS}}
\def \rosat{{\it ROSAT}}
\def \einstein{{\it Einstein}}
\def \euve{{\it EUVE}}
\def \orfeus{{\it ORFEUS}}
\def \chandra{{\it Chandra}}
\def\lesssim{\mathrel{\hbox{\rlap{\hbox{\lower4pt\hbox{$\sim$}}}\hbox{$<$}}}}
\def\gtrsim{\mathrel{\hbox{\rlap{\hbox{\lower4pt\hbox{$\sim$}}}\hbox{$>$}}}}

\def\ion#1#2{#1$\;${\small\rm#2}\relax}

\title{ EUV/X-ray Emission and the Thermal and Ionization Structure of B Star 
Winds}

\author{David H. Cohen}
\affil{Bartol Research Institute, University of Delaware, Newark, DE 19716; 
Prism Computational Sciences, 16 N. Carroll St., Ste. 950, Madison, WI 53703}

\author{Joseph P. Cassinelli}
\affil{Department of Astronomy, University of Wisconsin, Madison, 475 N. Charter St., Madison, WI 
53706}

\author{Joseph J. MacFarlane}
\affil{Prism Computational Sciences, 16 N. Carroll St., Ste. 950, Madison, WI 
53703}

\author{Stanley P. Owocki}
\affil{Bartol Research Institute, University of Delaware, Newark, DE 19716}

\begin{abstract}

We discuss the EUV and X-ray properties of B stars, focusing on \ecma\ (B2 II) which is the {\it only} star with both emission lines and a photospheric continuum detected with \euve.  We explore the modest effects of the photospheric EUV continua on the wind, as well as the much stronger effects of the short-wavelength EUV and soft X-ray emission lines.  Attenuation of the EUV and soft X-ray emission by the wind plays an important role, and leads to the reprocessing of X-rays via He$^+$ ionization and the Bowen mechanism in the wind. Finally, we explore some of the new diagnostics that will shortly become available with the next generation of high spectral resolution X-ray telescopes. All of this analysis is presented in the context of a two component stellar wind--a dense component (clumps) that contains most of the mass but fills a negligible fraction of the volume, and a rarefied component that fills most of the volume but accounts for only a small fraction of the mass. 

\end{abstract}


\section{Introduction}

The primary differences between the winds of B stars and those of O stars are the much lower densities of B star winds and the relative importance of 
EUV/X-ray emission in B stars.  In this paper we will examine the ionization and temperature structure of B star winds, with a focus on the EUV and X-ray emission properties of these objects.  We will see how the differences in the wind and EUV/X-ray properties cause strong departures from the thermal and ionization conditions found in the more well-studied and 
well-understood O stars. The ionization balance in B star winds is both more complex, and harder to determine, both theoretically and observationally, than in O star winds.  In fact, because of the more uncertain ionization corrections to the UV mass-loss rates the actual mass-loss rates of B stars are not well known.  

In addition to this crucial difference between the high wind densities of O stars and the low wind densities of B stars, the relative proximity of the nearest B stars provides us with some important information that does not exist for O stars.  Due to the very low column densities toward $\epsilon$ CMa and $\beta$ CMa, we have information about the extreme ultraviolet (EUV) emission properties--both of the photosphere and of the wind--of these two objects, while we have no comparable information about O stars.  As we shall see, this EUV emission plays an important role in the setting the ionization conditions (and possibly the thermal conditions) in the low-density B star winds.

The dense winds of O stars give rise to ultraviolet absorption lines that show both P Cygni emission peaks and broad absorption troughs.  However, in B stars the winds are not strong enough to generate emission components, and the observed blue-shifted absorption components are seen to be quite weak in the few ions that are observable in the spacecraft ultraviolet. 
Grigsby \& Morrison (1995) have shown that UV wind features (at least in those ions accessible to IUE) do not exist beyond about B2.  


In principle, radiation-driven winds can exist if there is at least one optically thick line in the wind.  Gayley (1995) has developed a new formalism for describing the radiation force in standard line-driven winds (Castor, Abbott, \& Klein 1975; hereafter CAK).   He has shown that the line opacity in a stellar wind has the remarkably simple character that its opacity is ${\bar Q} \approx 2000$ times stronger than free electron scattering. This includes the effects of elemental abundances and assumes a matching between the wavelengths at which the line opacity exists and those at which the photospheric spectrum is strongest.  Therefore, stars with Eddington factors $\Gamma > 10^{-3}$ have the potential for line-driven winds.  However, Puls (2000) has recently shown that for B stars, the opacity and photospheric flux may not be well matched, making it harder for B stars to drive a strong outflow.   In any case, this $\Gamma > 10^{-3}$ limit corresponds to stars that have somewhat later spectral types than the B2 limit seen in the UV wind lines.  The disappearance of these spectral wind features may, however, be due to ionization effects, rather than the disappearance of winds altogether. As we have already pointed out, determining the mass-loss rates of B stars is very difficult because the ionization corrections to UV absorption line observations is so uncertain, and no other reliable mass-loss diagnostics exist for most of these stars. 

Despite the findings of Grigsby and Morrison (1995), there is, in fact, some evidence for winds in mid- to late-B, and even early-A, stars.  Much of the evidence is indirect and centers on the X-ray emission (which we will discuss in \S3), but one intriguing piece of more direct evidence involves the detection of a blue-shifted absorption component in the 
Lyman-$\alpha$ line of the A0 V star Sirius (Bertin et al. 1995). This feature is consistent with a wind having a terminal velocity of $\vinf \approx 100$ km s$^{-1}$ and a mass-loss rate of $\mdot \approx 10^{-12}$ \msunyr.  So, it seems that at least some stars with $\Gamma < 10^{-3}$ have winds, although they may {\it not} be driven (primarily) by radiation pressure acting on lines.

In the remainder of this paper we discuss some results from large surveys but concentrate on detailed observations of the B2 II MK standard $\epsilon$ CMa.  With a wind-broadened ($v_{\rm edge} \approx 800$ km s$^{-1}$), but weak, \ion{C}{IV} 1548, 1551 \AA\/ profile and a mass-loss rate of $\mdot \approx 1$ to $2 \times 10^{-8}$ \msunyr\ (Cassinelli et al. 1995; Cohen et al. 1999) this star has a much lower density wind than O stars, but, unlike some later B stars, the existence of its wind is not in doubt.  We note that with $\Gamma \approx 0.01$ the multiplier factor of ${\bar Q} \approx 2000$ implies that the force from spectral lines exceeds gravity at the star's surface.  This star has $\teff = 21000$ K and $\lxlb = 10^{-7}$.  It
is very well studied and, most importantly, is the only hot star observable with all three of the Extreme Ultraviolet Explorer (\euve) spectrometers.  This unique data set alone argues for concentrating on \ecma\/ when trying to understand 
radiation-driven winds in B stars.



\section{Photospheric EUV Emission from B Stars and its Effect on the Wind}

Direct measurements of stellar photospheric continua below the Lyman edge at 912 \AA\/ were impossible until the early 1990s when the \euve\/ satellite was launched.  Even with this instrument, only a small number of stars could be observed, due to a combination of low intrinsic fluxes and high interstellar opacity.  Of these stars (aside from white dwarfs), all of the observations are of coronal emission lines from late-type stars, except for two stars: \ecma\/ and \bcma.  These are the only two stars for which there are direct measurements of the photospheric flux below the Lyman edge.  This is due primarily to their fortuitous position in a tunnel of very low interstellar column density. 

In general, any study that has required the photospheric EUV fluxes of hot stars--including studies of the warm ionized component of the interstellar medium as well as studies of stellar wind conditions--has relied on model atmospheres.  There are also a very small number of {\it indirect} observations of EUV fluxes of B stars.  One of these involves Fabry-Perot spectroscopy of H$\alpha$ fluxes of small \ion{H}{II} regions around B stars.  Assuming that all stellar Lyman continuum radiation is reprocessed into H$\alpha$ in these \ion{H}{II} regions, Kutyrev, Reynolds, \& Tufte (1993) and Reynolds (1995) placed lower limits to the stellar Lyman continuum fluxes that in some cases exceed the predictions of model atmospheres. 

The direct measurements of the Lyman continuum fluxes of \ecma\/ and \bcma\ (B1 III) with \euve\/ revealed fluxes that are 10 to 100 times higher than the predictions of model atmospheres (Vallerga, Vedder, \& Welsh 1993; Cassinelli et al. 1995; Cassinelli et al. 1996). The discrepancy seems to be even higher in the \ion{He}{I} continuum ($\lambda < 504$ \AA), which is observed in \ecma\/ but not \bcma, although the correction for interstellar attenuation is more uncertain at these wavelengths.  Non-LTE, line-blanketed TLUSTY models (Hubeny \& Lanz 1995) actually provide worse fits to the data, as the additional physical effects tend to both cool the outer wind, and lead to an overpopulation of the ground state with respect to LTE.  Both these effects lower the Lyman continuum flux.


Cassinelli et al. (1995; 1996) suggest that whatever the physical mechanism, the EUV excesses and the observed IR excesses can both be accounted for by a modest increase in the temperature with respect to models in the extreme outer layers of the atmosphere ($\tau_{\rm Ross} \approx 0.01$), perhaps of 1000 to 2000 K.  The IR continuum is dominated by free-free opacity, and as such, is in LTE.   Furthermore, the IR continuum near 15 \microns\/ is formed in the {\it same physical layers} as the EUV continuum near 600 \AA.  The 1000 K to 2000 K temperature increase that can explain the observed Lyman continuum excess in \ecma\/ and \bcma\/ can also explain the roughly 10\% IR excess observed with \iras.  Although non-LTE effects could in principle explain the observed EUV excess, only the existence of a temperature above the predictions of models high in the atmosphere can {\it simultaneously} explain the EUV and IR excesses. 

\begin{figure}
\plotfiddle{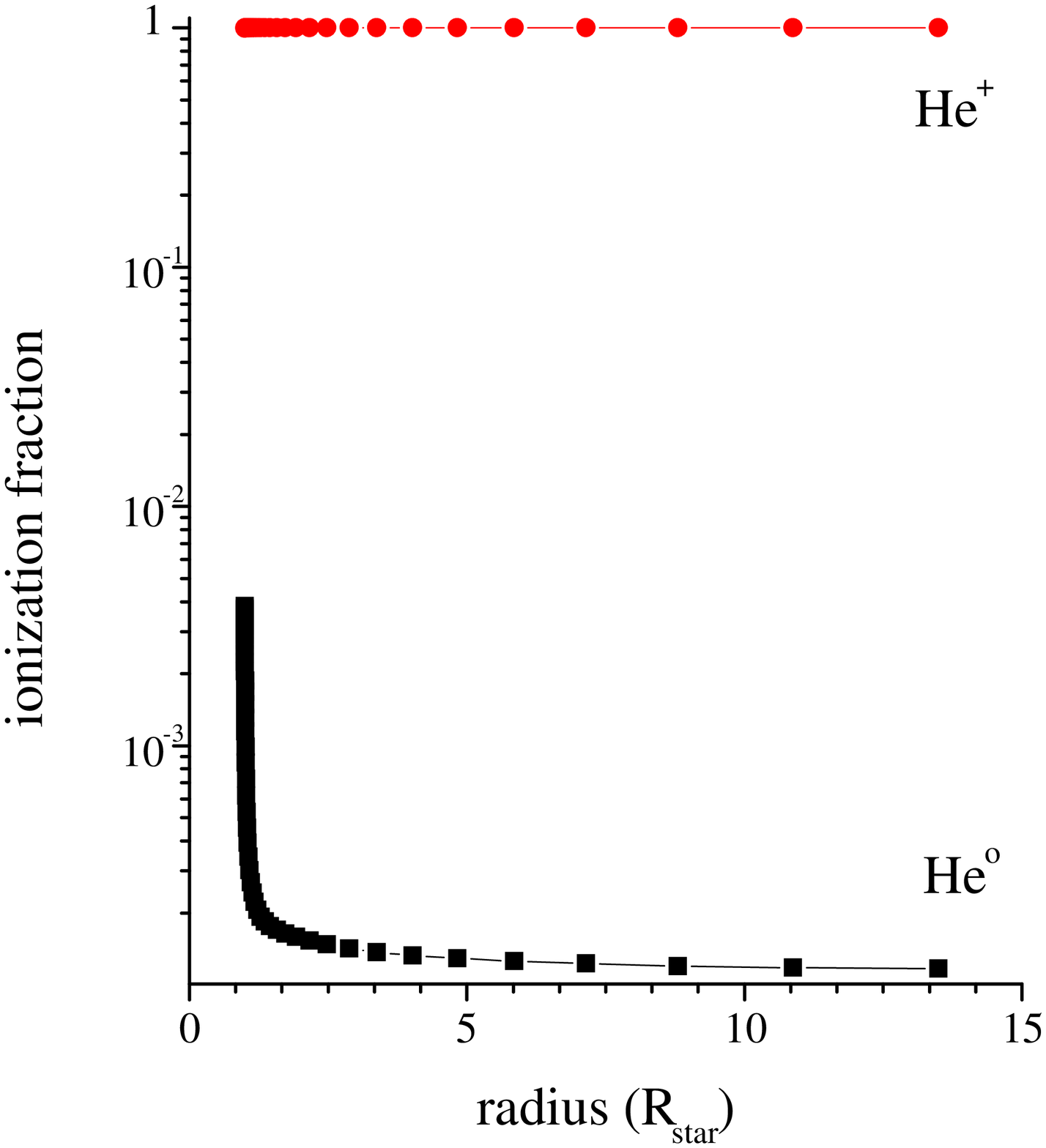}{2.5in}{0}{30}{23}{-170}{59}
\plotfiddle{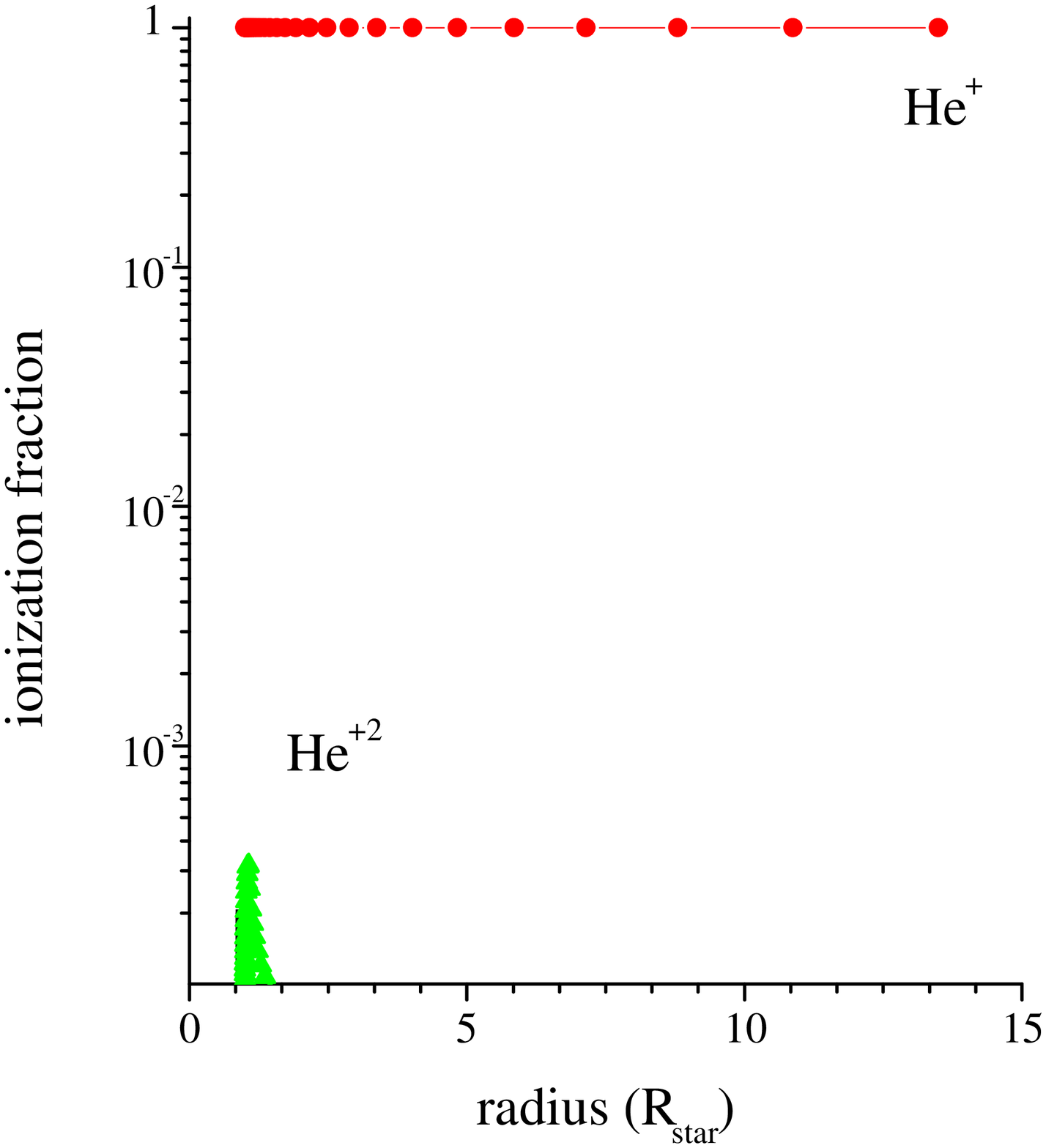}{0in}{0}{30}{23}{-170}{-68}
\plotfiddle{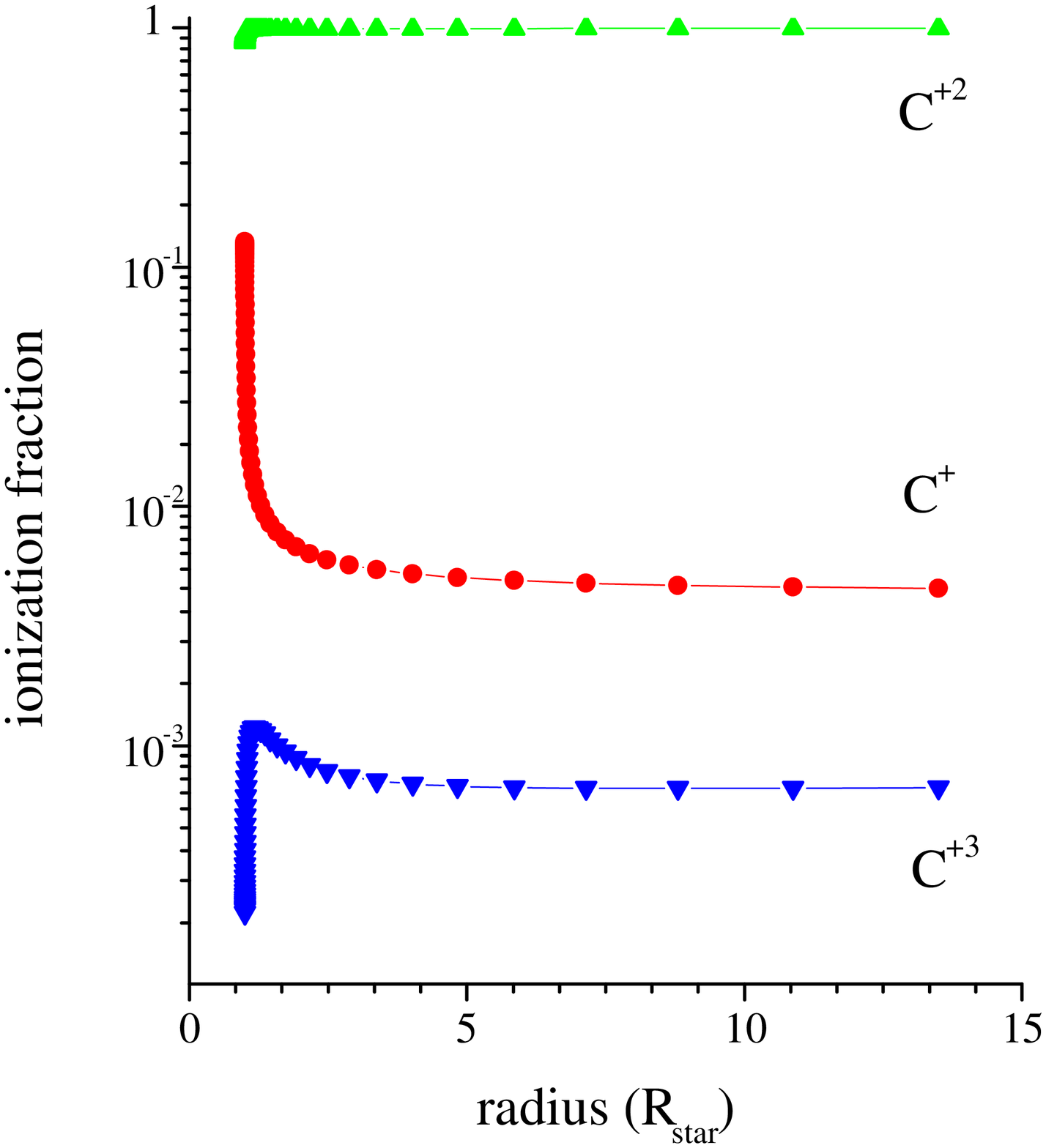}{0in}{0}{30}{23}{0}{107}
\plotfiddle{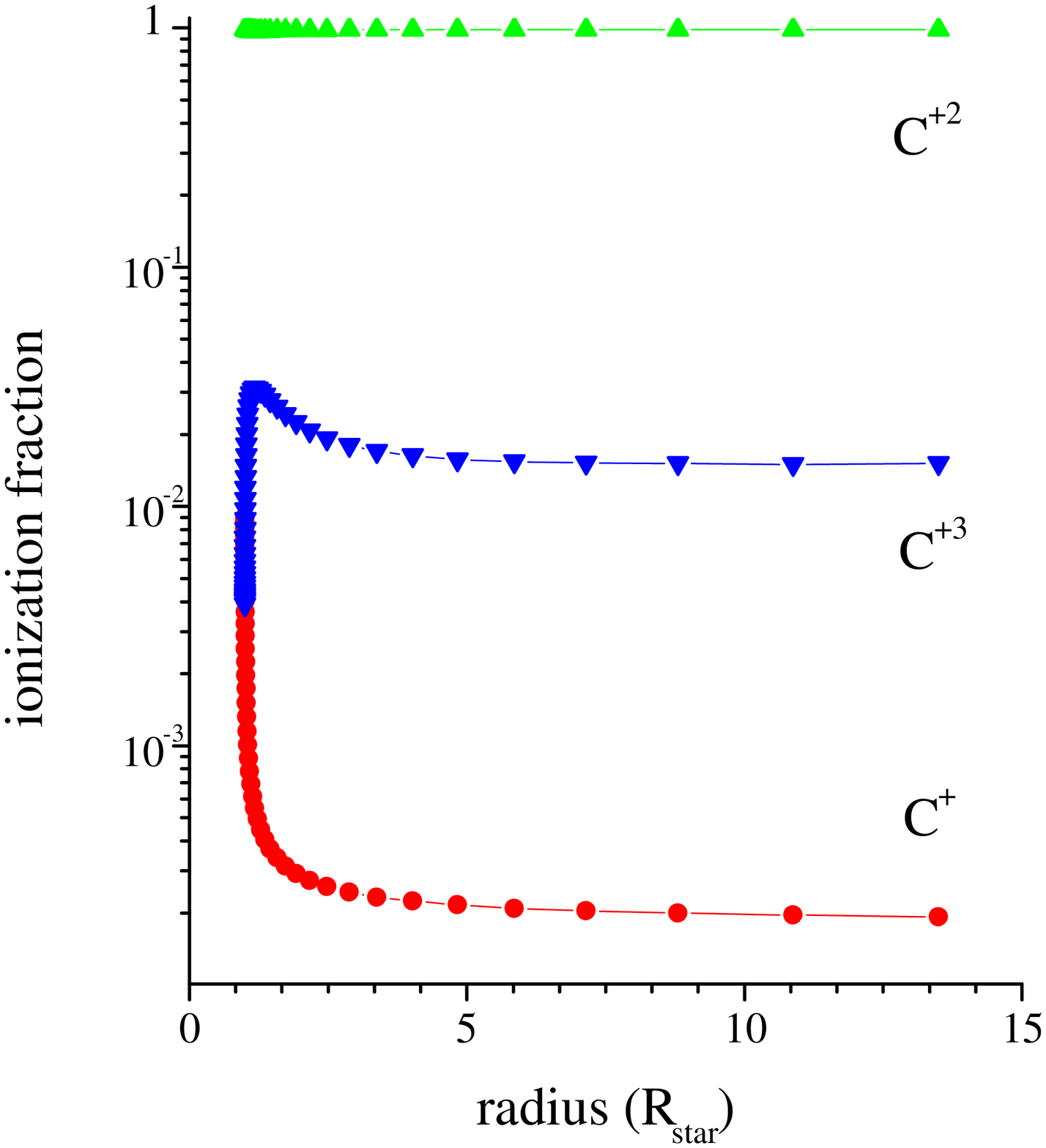}{0in}{0}{30}{23}{0}{-20}
\caption{Ionization balance for helium (left) and carbon (right) with two different photospheric flux levels: the observed high levels (bottom) and the predicted lower levels (top).} 
\label{Fig:photospheric_ionization}
\end{figure}

It is not clear how representative these two stars are, or how large the region of the HR diagram is for which B star EUV fluxes have been underestimated.  However, the indirect evidence of \ion{H}{II} region with high H$\alpha$ fluxes indicates that these stars are not unique.  To investigate the influence of the photospheric EUV flux levels of B stars on the wind ionization, we have calculated non-LTE ionization equilibrium models for the wind of \ecma\/ using both the observed spectral energy distribution, with its high EUV flux, and a spectrum consistent with non-LTE models, with its much lower EUV flux.  These calculations are identical to the ones described in MacFarlane, Cohen, \& Wang (1994), where the authors used a version of the NLTERT code (MacFarlane 1997) that accounts for spherical expansion.  In this code, rate equations for multi-level atoms are solved self-consistently with the radiation field.  We show a comparison in Figure \ref{Fig:photospheric_ionization} of the helium and carbon ionization balance as a function of radius for the two different cases.  The effect of the factor of 100 higher EUV flux is not major, but is nonetheless significant.  In both cases, He$^+$ and C$^{+2}$ dominate, but the higher photospheric flux changes the second most abundant ionization state to the one {\it above} the dominant stage from the one {\it below} it. This does not change the bulk properties of the wind, but it can have a strong impact on the strengths of the observed UV resonance lines.  Specifically, note that the mass-loss rate determined from the \ion{C}{IV} 1548, 1551 \AA\/ doublet will be roughly a factor of 20 higher if the ionization correction fails to account for X-ray photoionization.

The effect of the higher EUV fluxes in B stars on the wind temperature has yet to be explored.  However, it is clear that the hydrogen ionization rate will be strongly increased by the higher fluxes, and photoionization is the primary heating mechanism in OB star winds (see Drew 1989).

\section{X-ray and Short-Wavelength Emission from B Stars and its Effect on the Wind}

The short-wavelength EUV ($40$ eV $\lesssim h{\nu} < 100$ eV) and soft X-ray line emission from early-B stars is usually understood in the context of the wind-shock emission that is assumed to operate in O stars.  The X-ray emission from mid- and late-B stars is harder to understand in terms of the standard picture of wind shocks, however. Much of the information we have about the X-ray properties of B stars comes from large low-resolution surveys with \rosat\/ and \einstein, but a significant amount of complementary information has been obtained from higher resolution observations of a small number of B stars, including \ecma.

The surveys have indicated that the $\lxlb \approx 10^{-7}$ relation that holds for O stars extends to about B1 but then breaks down, with $L_{\rm X}$ decreasing faster than $L_{\rm Bol}$ beyond this spectral subtype (Cohen, Cassinelli, \& MacFarlane 1997). The breakdown in the \lxlb\/ law occurs just at the point where hot star winds are becoming optically thin to X-rays, as discussed by Owocki \& Cohen (1999), who show that the X-ray luminosity is expected to scale as $L_{\rm X} \sim \mdot^2 \sim L_{\rm Bol}^{3.4}$.  However, the slope in the \lxlb\/ relationship is not observed to be this steep, implying that the X-ray filling factors in mid-B stars approach, and in some cases exceed, unity (Cohen et al. 1997).
There are several possible implications to these surprising data: (1) the X-ray emitting gas is clumped to enhance the emission measure for a given total mass; (2) the mass-loss rates of these B stars are actually higher than Abbott's (1982) CAK parameters imply; or (3) the X-rays do not arise (entirely) in the wind. Note that this filling-factor problem is not as significant an issue for very early B stars and early-B stars lying above the main sequence, like \ecma, which exhibits an emission measure filling factor of roughly 0.1 (Cohen et al. 1996). 

\begin{figure}
\plotfiddle{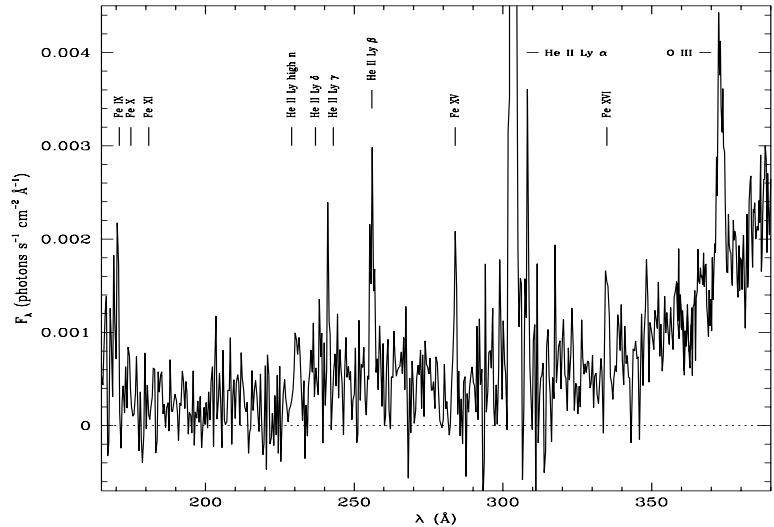}{2.8in}{0}{55}{38}{-175}{-72}
\caption{The \euve\/ spectrum of \ecma.  Note the photospheric continuum at the longer wavelengths, as well as low ionization emission lines (\ion{He}{II} and \ion{O}{III}) and high ionization lines of various stages of iron.} 
\label{Fig:euve_data}
\end{figure}

The \euve\/ spectrometers have access not only to photospheric radiation in the so-called ``long-wavelength'' spectrometer ($300$ \AA $\lesssim \lambda \lesssim 760$ \AA), but also to line emission in the short- and medium-wavelength spectrometers ($70$ \AA\ $< \lambda < 370$ \AA).   For \ecma, five highly ionized iron lines are detected in the \euve\/ data with greater than $4 \sigma$ significance.  These lines, along with emission lines formed in the photoionized cool component of the stellar wind and the \ion{He}{I} photospheric continuum, are shown in Figure \ref{Fig:euve_data}.  
These iron lines have a combined {\it emergent} luminosity on par with the total X-ray luminosity observed with \rosat, and significantly more {\it intrinsic} luminosity (the difference being the result of continuum opacity in the wind).  They can be fit simultaneously with the \rosat\/ data to derive a temperature distribution of the hot plasma on \ecma.  The results of this modeling indicate that there is significantly more plasma with temperatures at or below $10^6$ K than plasma with temperatures near $5 \times 10^6$ K (Cohen et al. 1996). This result is consistent with the \rosat\/ data for other B stars (Cohen et al. 1997).  The analysis of the combined \euve\/ and \rosat\/ data also indicates that the wind is somewhat optically thick at very soft X-ray energies, and quite optically thick at EUV energies.  The amount of attenuation derived from these data indicates that the hot plasma on \ecma\/ is spatially distributed throughout the wind (Cohen et al. 1996). 

\begin{figure}
\plotfiddle{He-high.eps}{2.55in}{0}{30}{23}{-170}{54}
\plotfiddle{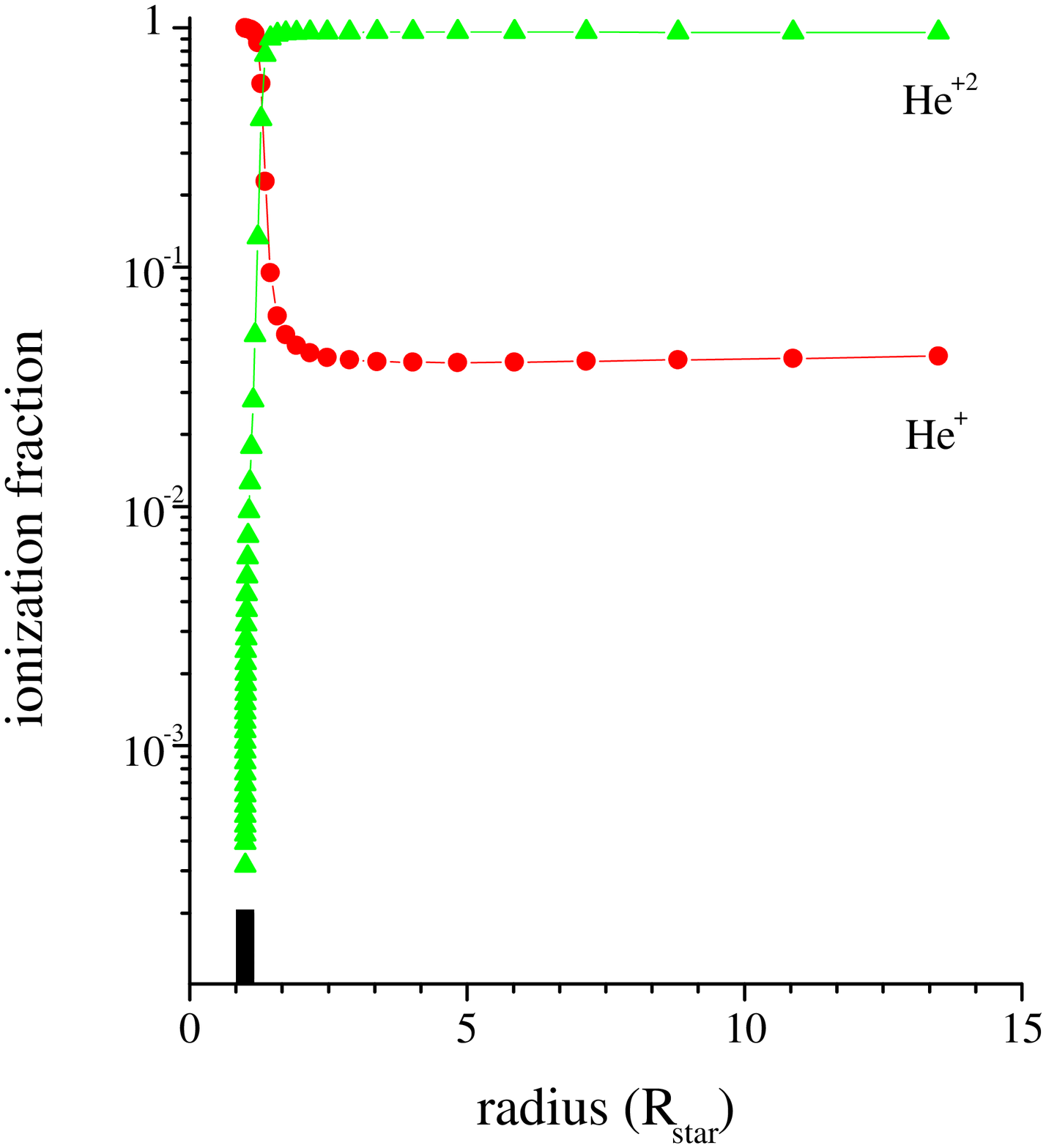}{0in}{0}{30}{23}{-170}{-68}
\plotfiddle{C-high.eps}{0in}{0}{30}{23}{0}{102}
\plotfiddle{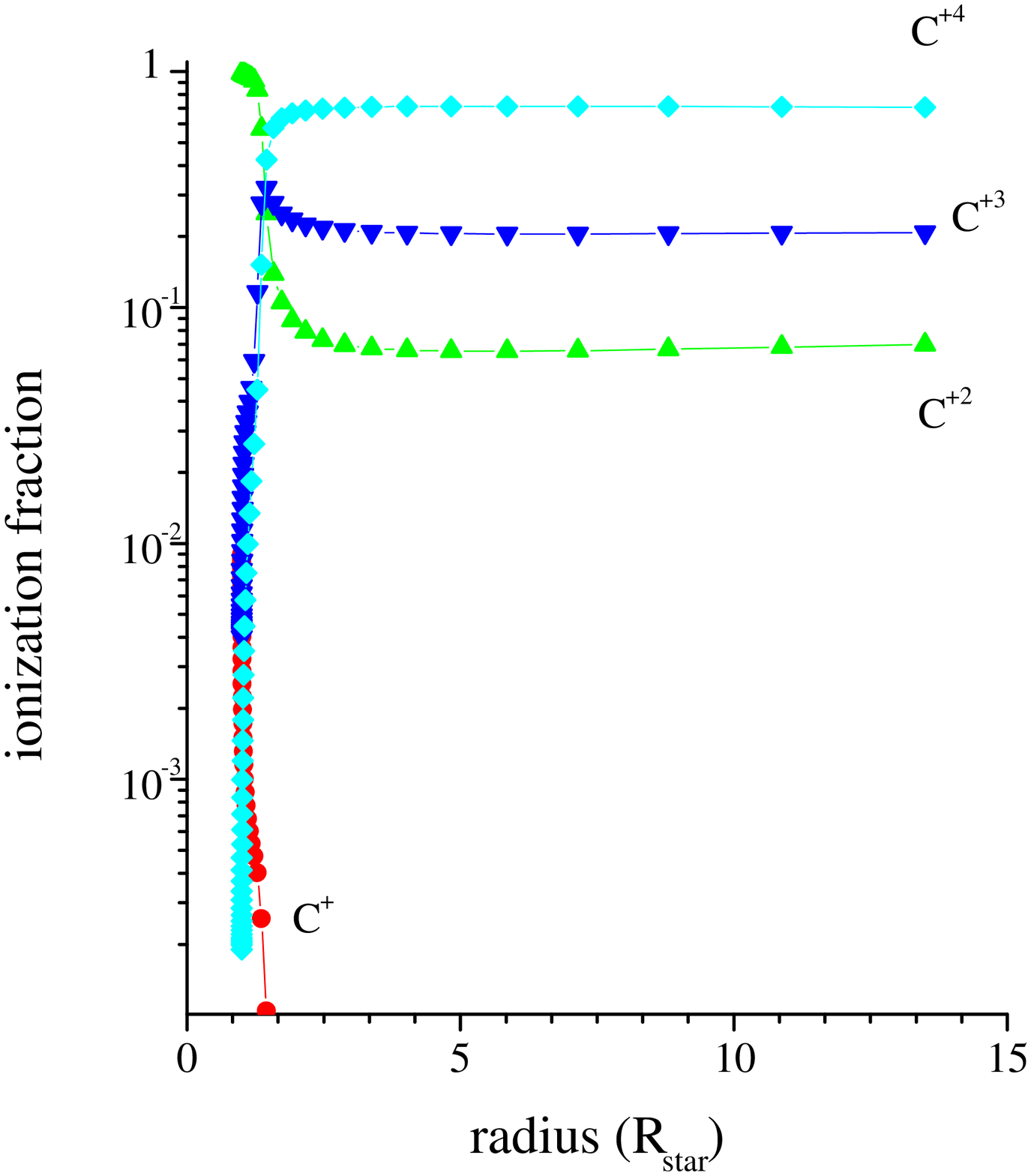}{0in}{0}{30}{23}{0}{-20}
\caption{Ionization balance for helium (left) and carbon (right) with X-rays (bottom) and without X-rays (top). 
} 
\label{Fig:xrays_ionization}
\end{figure}

The information about the temperature distribution of the X-ray emitting plasma on \ecma, as well as the wind absorption, allow us to generate and constrain models of the wind ionization state of the bulk, unshocked wind.
In much the same way we compared ionization state models for different values of the photospheric EUV flux, let us now examine the effect of X-rays on the wind ionization.  Using an input photospheric spectrum for which we have adjusted the EUV continuum flux to match the \ecma\/ data, we calculated a model with an X-ray source distributed throughout the wind and compared it to the model with no X-rays. In Figure \ref{Fig:xrays_ionization} we show the results of this comparison for helium and carbon.  
It can immediately be seen that the X-rays' influence is much greater than the photospheric EUV radiation's.  With the X-ray photoionization in the calculation, the primary ionization state changes from He$^+$ to He$^{+2}$ and from C$^{+2}$ to C$^{+4}$.  There is also a general spreading-out of the ionization stages, with three different stages of carbon each accounting for at least 10\% of the total.

An explanation for the strong effect of X-rays in determining the ionization states of B star winds (as opposed to O stars winds in which the photosphere sets the ionization rate, and the X-rays cause only a perturbation) has been offered by MacFarlane et al. (1994). In the hotter O stars, the photospheric radiation field is strong above the \ion{He}{II} edge at 228 \AA\/ (54.4 eV) and the X-ray field dominates the photosphere only above about 100 eV. By contrast, for the cooler B stars, the X-ray field dominates the much weaker \ion{He}{II} continuum and even some of the \ion{He}{I} continuum, starting at about 40 eV.  This causes the X-rays to have a stronger effect on the wind ionization in B stars.  In fact, the X-rays boost the mean ionization state about one full stage on average for B stars.  

\begin{figure}
\plotfiddle{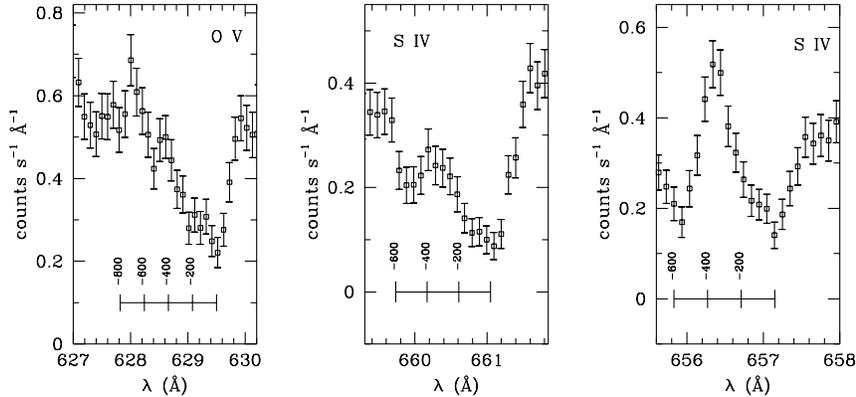}{2.in}{0}{60}{60}{-182}{-270}
\caption{Wind-broadened EUV features seen in the {\it ORFEUS-SPAS} spectrum of \ecma.}
\label{Fig:orfeus}
\end{figure}

Indeed, in \ecma\/ surprisingly high levels of wind ionization are seen in oxygen.  Both \euve\/ and \orfeus\/ detected wind broadening in the \ion{O}{V} 630 \AA\/ line.  See Figure \ref{Fig:orfeus} for the higher-resolution \orfeus\/ spectral measurements.  {\it Both} X-rays and a high photospheric EUV flux are  necessary to produce significant amounts of O$^{+4}$ in the wind of this star. 

\section{The Bowen Mechanism in the Wind of \ecma}

\begin{figure}
\plotfiddle{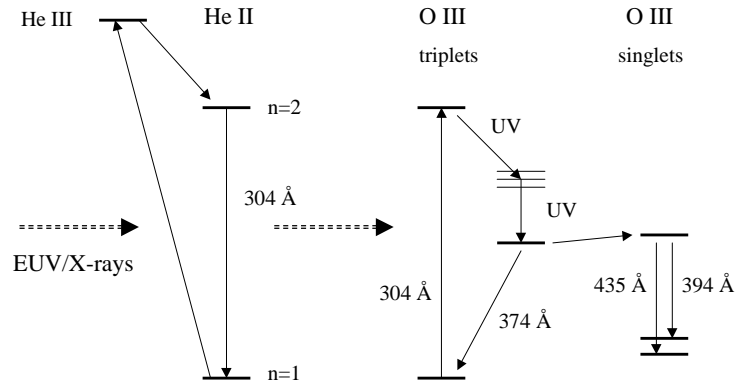}{2.1in}{0}{50}{50}{-145}{-125}
\caption{Schematic energy-level diagram of the Bowen fluorescence mechanism.  In \ecma, and likely in other hot stars, the X-rays that arise in the wind ionized helium, which upon recombination populates the He$^+$ $n=2$ state, which leads to Ly$\alpha$ emission at 304 \AA.  This excites an \ion{O}{III} transition out of the ground state.  The subsequent cascade produces 374 \AA, 394 \AA, or 435 \AA\/ emission, with a branching ratio that is temperature sensitive.}
\label{Fig:grotrian}
\end{figure}

It should be noted that the attenuation of X-rays in the bulk stellar wind involves primarily very soft X-rays and short-wavelength EUV emission, and that the majority of photoionization involves helium (from singly to fully ionized).  The ionization of helium by the X-rays leads to recombination radiation, primarily in the \ion{He}{II} Ly$\alpha$ line at 304 \AA.  This line is coincident with an \ion{O}{III} line, which is pumped, leading to a radiative cascade that (usually) culminates in the emission of a photon at 374 \AA.  Until it was seen in \ecma, this line had never been detected in an astrophysical object outside the solar system, although the near-UV lines emitted by transitions among the high-lying levels involved in the Bowen mechanism are often seen in nebulae. In Figure \ref{Fig:grotrian} we show an energy level diagram for the Bowen mechanism in \ecma. 

Cohen et al. (1999) have shown that the 304 \AA\/ and 374 \AA\/ fluxes 
are quite sensitive to the mass-loss rate.  They find a value of $\mdot \approx 1$ to $2 \times 10^{-8}$ \msunyr, which is about a factor of three less than the prediction from the CAK parameters compiled by Abbott (1982). The 304 \AA\/ line is wind broadened in the \euve\/ spectrum of \ecma.  The broadening is consistent with a line that is optically thick out to velocities of $v = 750 \pm 150$ km s$^{-1}$. This is further indication that the X-ray emission is distributed throughout the wind, and furthermore, that it interacts with the bulk wind and causes ionization throughout the wind.  

Finally, it has recently been realized that the Bowen fluorescence can resolve via one of two transitions on the singlet side of \ion{O}{III} at 394 \AA\/ and 435 \AA, in addition to the 374 \AA\/ triplet (Cohen et al. 1999).  The relative strength of the singlet and triplet lines is sensitive ultimately to the wind temperature, and so these lines may prove to be the most quantitative and direct temperature diagnostic of any hot star wind. 


\section{Upcoming X-ray Spectroscopy Missions}

An entirely new era of astrophysical X-ray spectroscopy is imminent with the recent launches of \chandra\/ and \xmm.  The spectral diagnostics that will soon become available will allow for the determination of the density, velocity, and temperature of the hot regions in the winds of B stars, and other hot stars.  We will be able to answer questions such as: Are the shocked regions denser than the ambient wind?  And what is the ionization distribution in the denser structures in B star winds?

In order to demonstrate the types of analysis that will be possible with these new data--and also to explore directly the clumping properties of B star 
winds--we briefly discuss some new interpretations of recent 
medium-resolution \asca\/ spectra of the X-ray binary, Vela X-1, which contains a B0.5 I star as the mass-losing companion. The accretion of the B star's wind onto the compact companion produces hard X-rays, which are reprocessed in the wind.  When viewed during eclipse, virtually all of the radiation is coming from the B star's wind, primarily via recombination in the highly (photo-)ionized portion of the wind and via X-ray fluorescence in the less ionized, denser portion of the wind. 

\begin{figure}
\plotfiddle{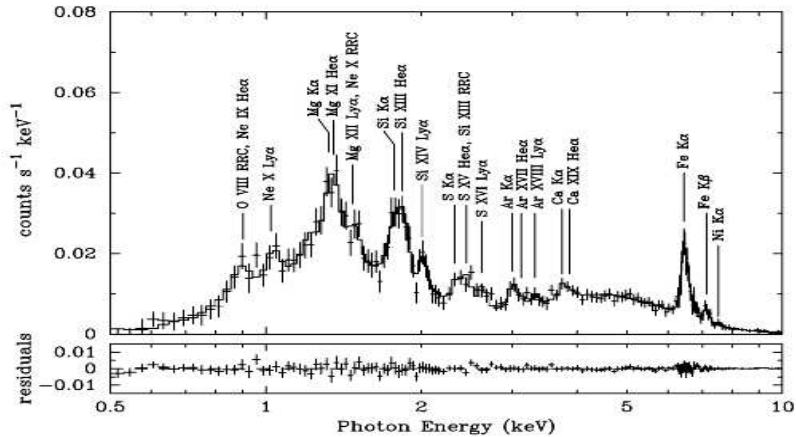}{2.2in}{0}{60}{45}{-152}{-12}
\caption{The \asca\/ spectrum of Vela X-1 shows emission lines that arise from recombination in the low-density component of the wind dominate the spectrum.  However, several features from the cold, dense wind component are seen.  These are the radiative recombination continua (RRC) and several fluorescent lines that are blended with emission features from highly ionized states of the same elements.}
\label{Fig:velax1}
\end{figure}

Recently, Sako et al. (1999) have identified these two different spectral components in the \asca\/ data (see Figure \ref{Fig:velax1}).  Much more quantitative analysis will be possible with the higher resolution spectra that become available over the next few years.  However, Sako et al. (1999) extract a significant amount of information from this marginal spectrum. They fit a mass-loss rate and velocity $\beta$ parameter for the B star's wind, based on observations of the highly ionized recombination lines which are formed in the hot, low density component.  They also derive the overall mass in the cool dense component and put some constraints on its density from the observations of the X-ray fluorescence lines.  The results of the analysis show the highly ionized low density component accounts for less than 10\% of the mass, but occupies 95\% of the volume.  Conversely, the cold, dense clumps occupy less than 5\% of the volume but represent about 90\% of the mass of the wind.  

It is notable that this type of wind structure is similar to that seen in numerical simulations of the line-force instability (Owocki, Castor, \& Rybicki 1988; Feldmeier 1995; Feldmeier, Puls, \& Pauldrach 1997; Owocki, Runacres, \& Cohen, these proc.).  Furthermore, to maintain clumping structure far out in the wind flow, over-dense and under-dense regions must be close to pressure equilibrium.  The ``ambient'' wind (the mass-weighted mean wind, which is dominated by the dense clumps) has a temperature somewhat below the stellar effective temperature ($T \sim 10^{4}$ K) and the X-ray emitting component has a typical temperature of $T \sim 10^6$ K.  Pressure equilibrium therefore requires the hot component to be of order 100 times less dense than the cool component.  Using the numbers from the \asca\/ observations of Vela X-1, the clumps are 20 times denser than the mean wind\footnote{In the two component picture, this ``mean wind'' exists only in a statistical sense.}, while the hot, low-density component is 10 times less dense.  So the density contrast between the two components is 200.  

Owocki \& Cohen (1999) have introduced a simple empirical model for X-ray transfer through a two-component wind in an effort to explain the $\lxlb \sim 10^{-7}$ law.  The total X-ray flux generated scales as the observed emission measure, 

\begin{equation}
L_X \sim (E.M.) \Lambda
\end{equation}
\noindent
where $\Lambda$ is a function of atomic parameters and of the ionization distribution in the emitting plasma, and the emission measure is given by

\begin{equation}
E.M. = \int n_{\rm e} n_{\rm i} dV .
\end{equation}
\noindent
Here the volume integral is over regions with plasma at X-ray emitting temperatures ($T \gtrsim 5 \times 10^{5}$).  Owocki \& Cohen (1999) scale this to the ``whole wind'' emission measure, as 
\begin{equation}
E.M. = C_{\rm s} f_{\rm m} (E.M._{\rm wind}) = C_{\rm s} f_{\rm m} \int^{\infty}_{R_{\rm min}} n_{\rm e} n_{\rm i} 4{\pi}r^2 dr
\end{equation}
\noindent
where the mass filling factor, $f_{\rm m}$, is the fraction of the mass of the wind (above some appropriate minimum radius) that is heated to X-ray emitting temperatures, and $C_{\rm s} = \rho_{\rm s} / {\bar \rho}$ is the typical density in the post-shock region compared to the local value of the density in the idealized smooth, mean wind. Note that each of these factors can be thought of as scaling one of the density terms inside the volume integral.

The product $C_{\rm s} f_{\rm m}$ is the normalization factor--essentially the emission measure filling factor--and the terms cannot be separated using 
low-resolution data;  but using spectroscopic techniques, enough information can be determined about both components of the wind to separate them.  For the B star in Vela X-1, $f_{\rm m} \approx 0.1$, and $C_{\rm s} \approx 0.1$.  This normalization factor $C_{\rm s} f_{\rm m} \approx 0.01$ is consistent with the \rosat\/ observations of O stars, and the B0.5 I component of Vela X-1 is a supergiant and so has a denser wind that is more like an O star's.

Thus, a consistent picture emerges of hot star winds, in which a pervasive hot, low-density component is in approximate pressure equilibrium with a cold, denser component presumably consisting of numerous small clumps.  The numbers are roughly consistent with numerical simulations of the line-force instability, perhaps having stochastic perturbations at the base (Feldmeier et al. 1997) for O and early-B stars.  They are also consistent with the overall levels of X-ray flux seen in O and early-B stars, as well as with the detailed X-ray reprocessing spectrum of Vela X-1, as observed with \asca\/ (Sako et al. 1999).  

For the mid- and late-B stars, however, recall that much higher filling factors ($C_{\rm s} f_{\rm m}$) are required to match the observations.  In these objects, the hot X-ray emitting gas must also be dense ({\it i.e.} $C_{\rm s} > 1$) to prevent $f_{\rm m} > 1$.  It is possible, for example, that magnetic loops confine a radiation-driven wind in some of these stars, forcing different flows to collide at high velocities, and containing the resulting hot gas, which preserves its relatively high density (Babel \& Montmerle 1997).  High-resolution X-ray spectroscopy will be able to provide density and temperature diagnostics of the emitting plasma in these stars, which will reveal whether the hot material in these winds exists in the dense phase.

\section{Summary}

The X-ray and EUV properties of B stars are significant, and have a controlling effect on the ionization conditions in the wind. With the improved understanding of these properties and their effects, additional modeling of the temperature structure in OB star winds is warranted. The stronger than expected photospheric EUV continua provide a small but important increase to the overall ionization conditions in the winds of B stars, but the X-ray and especially the short-wavelength EUV emission lines play the dominant role.  The ionization balance is changed completely by X-ray photoionization.  Furthermore, the X-rays ($h\nu > 100$ eV) that are seen by instruments like \rosat\/ represent just a small fraction of the photons emitted by the hot plasma in the winds of B stars, which are predominantly EUV photons. 

Finally, the interaction of this EUV/X-ray emission with the cooler, denser wind component involves a complex reprocessing of the absorbed X-ray photons into UV and EUV emission lines via the Bowen fluorescence mechanism.  More and more, the detailed spectral information that is becoming available will drive our understanding of the structure of hot star winds, including those of B stars.  The picture that has started to emerge is one of a clumped, two-component medium, in which most of the bulk properties of the wind are defined by the dense clumps--in a sense, these clumps {\it are} the wind--but most of the volume is filled by a much lower density component.

The information we have covered leads us to identify the following questions which still need to be addressed:  What role does clumping play in B star winds, and how do clumps originate and evolve? How does X-ray induced ionization feed back into the line force, wind dynamics, and mass-loss rate?  How are thermal conditions affected by the high photospheric EUV fluxes?  And what is the nature of winds for stars later than about B2?

\acknowledgments

We thank Mark Runacres for his careful reading of the manuscript.  This work 
was partially supported by NASA grant NAG 5-3530 to the Bartol Research 
Institute at the University of Delaware.

\end{document}